\documentclass[prd,preprint,superscriptaddress,preprintnumbers,byrevtex,showpacs,nofootinbib]{revtex4}
\usepackage{epsfig}
\usepackage{isolatin1}
\newcommand{\postscript}[2]{\setlength{\epsfxsize}{#2\hsize}
   \centerline{\epsfbox{#1}}}
\def\eslt{E_T^{\rm miss}}
\def\to{\rightarrow}

\def\tst{\tilde t}
\def\ttau{\tilde \tau}

\def\tg{\tilde g}

\def\tw{\widetilde W}
\def\tz{\widetilde Z}

\begin{document}

\title{Using $b$-tagging to enhance the SUSY reach of the CERN Large
  Hadron Collider }
\author{P.~G. Mercadante}
\email{mercadan@fnal.gov}
\affiliation{Instituto de F\'{\i}sica Te\'orica, 
Universidade Estadual Paulista, S\~ao Paulo -- SP, Brazil.}

\author{J.~K. Mizukoshi}
\email{mizuka@fma.if.usp.br}
\affiliation{Instituto de F\'{\i}sica, 
Universidade de S\~ao Paulo, S\~ao Paulo -- SP, Brazil.}

\author{Xerxes Tata}
\email{tata@phys.hawaii.edu}
\affiliation{Department of Physics and Astronomy, University of Hawaii,
  Honolulu, HI 96822}
%

\begin{abstract}

Assuming that supersymmetry is realized with parameters in the
hyperbolic branch/focus point (HB/FP) region of the minimal supergravity
(mSUGRA) model, we show that by searching for multijet + $\eslt$ events
with tagged $b$ jets the reach of experiments at the LHC may be
extended by as much as 20\% from current projections. The reason for
this is that gluino decays to third generation quarks are enhanced
because the lightest neutralino has substantial higgsino
components. Although we were motivated to perform this analysis because
the HB/FP region is compatible with the recent determination of the
relic density of cold dark matter, our considerations may well have a
wider applicability since decays of gluinos to third generation quarks
are favoured in a wide variety of models.

\end{abstract}



\preprint{
\preprint{IFT-P.017/2005, IFUSP-1612/2005, UH-511-1072-05}}

\bigskip

\maketitle


The WMAP collaboration \cite{wmap} has determined the cosmological
density of cold dark matter (CDM) to be,
\begin{equation}
\Omega_{CDM}h^2 = 0.1126^{+0.008}_{-0.009}\;. 
\label{wmap}
\end{equation}
This measured value 
is of the same order of magnitude as the density expected from the
production in the Big Bang of a stable, weakly interacting particle with
a mass ${\cal O}$(100~GeV), assuming only that it was in thermal
equilibrium at some point in the past \cite{dodd}. Hence, the precise WMAP
measurement provides a stringent constraint on all models that include
heavy, stable weakly interacting particles. However, since
the dark matter may well be made of several components, strictly
speaking (\ref{wmap}) only implies an upper bound on the density of such
particles.  In particular this bound applies to the stable lightest
supersymmetric particle (LSP), frequently the lightest neutralino
$\tz_1$, of $R$-parity conserving supersymmetric models that have been
the focus of much attention during the last twenty-five years
\cite{rev}, and leads us to conclude that,
\begin{equation}
\Omega_{\tz_1}h^2 \leq 0.129 \ \ (2\sigma)\;.
\label{bound}
\end{equation}
Compatibility with (\ref{bound}) is possible only if the neutralinos can
annihilate efficiently which, in turn, is possible only if one of the
following holds:
\renewcommand{\theenumi}{\roman{enumi}}
\begin{enumerate}
\item The lightest neutralino is hypercharge gaugino-like and
  annihilates via $t$-channel exchanges of relatively light ($\sim
  300$~GeV) sfermions \cite{bulk}. \label{item:i}

\item The neutralino mass $m_{\tz_1} \simeq {1\over 2}m_{A,H}$ so that
  it annihilates resonantly via the exchange of the neutral Higgs bosons
  $A$ or $H$ in the $s$-channel \cite{funnel}. Since these heavier Higgs
  bosons are typically quite broad, and because the neutralinos have
  thermal motions, resonant Higgs annihilation occurs over a rather
  large range of parameters. There is also a small range of parameters
  where resonant annihilation via the lightest scalar Higgs boson $h$
  leads to efficient neutralino annihilation \cite{lhiggs}. \label{item:ii}
  
\item The neutralino mass is close to that of a charged or coloured
  particle; since this latter particle can annihilate efficiently, the
  neutralino density is correspondingly reduced as long as its interactions
  maintain it in thermal equilibrium with the co-annihilating charged
  particle \cite{coannih}. \label{item:iii}

\item The parameter $|\mu|$ is small compared to the gaugino masses so
 that the lightest neutralino has significant higgsino components and
 annihilates effectively via couplings to electroweak gauge bosons 
\cite{matchevfp}. \label{item:iv}

\item The neutralino has significant $SU(2)$ gaugino components, and so
  annihilates to $W^+W^-$ via large isotriplet $SU(2)$
  couplings. In this case, the lighter chargino tends to be close in
  mass to $\tz_1$, so co-annihilations may also be important \cite{hansen}.
\label{item:v}
\end{enumerate}

Most SUSY analyses of the implications of the WMAP measurement have been
performed within the framework of the mSUGRA model \cite{msugra} which,
assuming radiative electroweak symmetry breaking,
is specified by the well known parameter set,
$$m_0, m_{1/2},A_0,\tan\beta,{\rm sign}(\mu).$$ Within this framework,
which has also been the paradigm for many phenomenological analyses of
supersymmetry (SUSY), neutralino annihilation as in item
\ref{item:i}.~occurs in the so-called \textit{ bulk region} with small
values of $m_0$ and $m_{1/2}$; as in \ref{item:ii}.~in the $A$ or $H$
funnels which occur only if $\tan\beta$ is large; as in
\ref{item:iii}.~only close to the boundary of parameter space where
$\ttau_1$ becomes the LSP \cite{stau}, or for special values of $A_0$,
where $\tst_1$ becomes the LSP \cite{stop}; and as in \ref{item:iv}.~in
the hyperbolic branch/focus point (HB/FP) region with large values of
$m_0$ and modest to large values of $m_{1/2}$ \cite{fppapers}. The last option
\ref{item:v}.~is not realized within mSUGRA or for that matter in any
model with unification of gaugino masses, or in any SUSY Grand Unified
Theory (GUT) unless the field that breaks SUSY also breaks the GUT
gauge symmetry \cite{anderson}.  In the recently studied
non-universal Higgs masses (NUHM) extensions of mSUGRA \cite{nuhm}, the
Higgs funnel occurs for any value of $\tan\beta$ while the low $|\mu|$
region is accessible even for relatively low values of $m_0$
\cite{ournuhm}.

Within the mSUGRA framework that we adopt for this study, it has been
shown that, with an integrated luminosity of 100~fb$^{-1}$, experiments
at the Large Hadron Collider (LHC) will probe all of the bulk region,
most of the Higgs funnel and, except for the largest values of
$\tan\beta$, all of the stau co-annihilation region allowed by the WMAP
data \cite{sasha}. The reach in the low $|\mu|$ HB/FP region, however,
cuts off around $m_{\tg}\sim 1.6-1.8$~TeV, where the signal from gluino
pair production becomes rate limited. Although $\tw_1$ and $\tz_2$ are
relatively light and will be abundantly produced at the LHC, if $|\mu|$
is indeed small the efficiency for the well studied trilepton signal
from $\tw_1\tz_2$ production \cite{trilep} is reduced, especially if we
are deep in the HB/FP region where the mass gap between $\tw_1$ (or
$\tz_2$) and $\tz_1$ becomes rather small and the daughter leptons as
well as $\eslt$ from charginos and neutralinos are soft.  In this case,
it has been shown \cite{tadas} that by implementing specially designed
cuts to separate the chargino pair production soft decay products from
Standard Model (SM) background, experiments at an $e^+e^-$ linear
collider operating at $\sqrt{s}=500-1000$~GeV will be able to probe
portions of the HB/FP region not accessible at the LHC. Since the LHC is
scheduled to commence operations in 2007, while a linear collider is
even very optimistically at least a decade away, it is clearly
worthwhile to explore all strategies that can potentially expand the LHC
reach, especially in this low $|\mu|$ region favoured by the WMAP
measurements.

An obvious option would be to
re-examine the trilepton signal to see whether it is possible to
separate it from SM background processes.\footnote{While we were
  preparing this manuscript, we learnt that the reach via the trilepton
  channel  has
  recently  been
  re-examined in Ref.~\cite{newbaer}, and found to be smaller or
  comparable to the reach in other 
  leptonic and the $\eslt$ channels even when $|\mu| <  M_{1,2}$.}
In this paper, however, we follow a completely different strategy and
focus on the signal from gluino pair production (since squarks are very
heavy). Our starting point is the observation that since the lighter
higgsino-like neutralinos and charginos couple much more strongly to the
third generation than to the first two generations, decays of gluino
into third generation fermions will be strongly enhanced so that the
signal may be expected to be rich in high $E_T$ $b$-jets \cite{dp}. In
contrast, the dominant SM backgrounds to multijet + $\eslt$ channels
which give the largest SUSY reach at the LHC come from $t\bar{t}$
production, from $V+{\rm jet}$ production ($V=W,Z$) and from QCD
processes. Since the latter two backgrounds are not expected to be
especially rich in hard bottom quark jets, and because experiments at
the LHC are expected to have good $b$ tagging capability, we explore
whether requiring the presence of tagged $b$-jets in the signal allows
us to probe portions of the hyperbolic branch that are inaccessible
using the by now standard analyses \cite{bcpt,atlas,cms} of the various
multijet + $\eslt$ channels at the LHC.  While $b$-tagging has been
suggested before to explore the nature of the underlying
model \cite{btag}, to our knowledge it has never been proposed as a tool
for SUSY discovery.  We remark that although the $b$-tagged signal will
 be rate limited, unlike the trilepton signal from $\tw_1\tz_2$
production, this signal will be relatively insensitive to how deep we
are in the HB/FP region.

We use the program ISAJET 7.69 \cite{isajet} with a toy calorimeter
described in Ref.~\cite{bcpt} for our analysis. Jets are found using a
cone algorithm with a cone size $\Delta R
=\sqrt{\Delta\eta^2+\Delta\phi^2}= 0.7$. Clusters with $E_T > 40$~GeV
and $|\eta({\rm jet})| <$ 3 are defined to be jets. Muons (electrons) are
classified as isolated if they have $E_T > 10$~GeV (20~GeV) and visible
activity in a cone with $\Delta R=0.3$ about the lepton direction
smaller than $E_T < 5$~GeV. We identify a hadronic cluster with $E_T \ge
40$~GeV and $|\eta(j)|< 1.5$ as a $b$ jet if it also has a $B$ hadron,
with $p_T(B) > 15$~GeV and $|\eta(B)| < 3$, within a cone with $\Delta R
= 0.5$ of the jet axis. We take the tagging efficiency $\epsilon_b=0.5$,
and assume that gluon and other quark jets can be rejected as $b$ jets
by a factor $R_b= 150$ (50) if $E_T < 100$~GeV ($E_T >
250$~GeV) and a linear interpolation in between \cite{brej}. While we
make no representation about the tagging efficiency and rejection against
light quark and gluon jets that will be finally achieved, especially at
high LHC luminosity and in the jetty environment of SUSY events, we felt
that an exploratory study of just how much $b$-jet tagging helps with the LHC
reach would be worthwhile.

Rather than perform time consuming scans over the WMAP favoured HP/FB regions
of the $m_0-m_{1/2}$ planes of the mSUGRA model for several values of
$\tan\beta$, we have chosen three diverse model lines for our
analysis. For each of these model lines, we take $\mu > 0$ (the sign
favoured by the result of experiment E821 at Brookhaven \cite{g-2}) and
fix $A_0=0$ --- our results are largely insensitive to this choice --- and
\begin{itemize}
\item $m_{1/2}= 0.295 m_0 - 477.5$~GeV with $\tan\beta=30$ for Model
Line {1},
\item $m_{1/2}= 0.295 m_0 - 401.25$~GeV with $\tan\beta=30$ for Model
Line {2}, and
\item $m_{1/2}= {17\over 60} m_0 - 390$~GeV with $\tan\beta=52$ for Model
Line {3}.
\end{itemize}
For values of $m_0 \agt 1500$~GeV, these model lines all lie in the WMAP
allowed HB/FP region of the mSUGRA parameter space delineated in
Ref.~\cite{sasha}. The first two model lines have an intermediate value
of $\tan\beta$ with Model Line {1} being deep in the HB/FP region
while Model Line {2} closer to the periphery of the corresponding
WMAP region. We choose Model Line {3} again deep in the HB/FP
region, but with a very large value of $\tan\beta$ to examine any
effects from  a very large bottom  quark Yukawa coupling. We take
$m_t=175$~GeV throughout this analysis. 

\begin{table}[tbh]
\begin{center}
\caption{The branching fractions for the decays of the gluino with
 $m_{\tg} \simeq 1650$~GeV to third generation quarks in the three WMAP
 allowed model lines introduced in the text, together with the
 corresponding values of the hypercharge gaugino mass parameter $M_1$,
 and the superpotential Higgs mass parameter $\mu$.}
\bigskip
\begin{tabular}{lccc}
\hline
\hline
&   Model Line {1} & Model Line {2}& Model Line {3} \\
\hline
$m_{\tg}$ (GeV) & 1637 & 1665 & 1665 \\
$\mu$ (GeV) & 181 & 335 & 164 \\
$M_1$ (GeV) & 269 & 278 & 275 \\ 
$\tg \to \tw_1^-t\bar{b}+\tw_1^+\bar{t}b$ & 37\% & 34\% & 39\% \\
$\tg \to \tw_2^-t\bar{b}+\tw_2^+\bar{t}b$ & 8\% & 11\% & 8\% \\
$\tg \to \tz_1t\bar{t}+\tz_1b\bar{b}$ & 15\% & 5\%& 17\% \\
$\tg \to \tz_2t\bar{t}+\tz_2b\bar{b}$ & 19\% &16\% & 20\% \\
$\tg \to \tz_3t\bar{t}+\tz_3b\bar{b}$ & 6\% & 17\% & 4\% \\
$\tg \to \tz_4t\bar{t}+\tz_4b\bar{b}$ & 4\% &5\% & 4\%  \\
\hline
\hline
\label{tab:decays}    
\end{tabular}
\end{center}
\end{table}

The branching fractions for the decays of the gluino with a mass $\sim
1650$~GeV
(close to the limit that can be probed at the LHC via the usual multijet
+ multilepton +$\eslt$ analyses) into third generation fermions  are shown
in Table~\ref{tab:decays} for the three model lines introduced above.
The following features of Table~\ref{tab:decays} are worth noting.
\renewcommand{\theenumi}{\arabic{enumi}}
\begin{enumerate}
\item In all three cases, almost 90\% of the gluino
decays are to the third generation, so that we expect very hard top and
bottom quark jets in SUSY events.
\item In Model Lines {1} and {3} that are deep in the HB/FP
  region, the gluino mainly decays to the higgsino-like lighter chargino and 
  the two lightest neutralinos; in Model Line {2}, $\tz_1$ is
  dominantly the hypercharge gaugino, so that gluino decays to $\tz_2$
  and to $\tz_3$ are favoured. 
\item The main difference due to the large $\tan\beta$ value for Model
  Line {3} is the increased branching ratio for the decays $\tg \to
  b\bar{b}\tz_i$ relative to $\tg \to t\bar{t}\tz_i$. Although we have
  not separated these out in the Table, we have checked that while the
  direct decays to bottom comprise just about 10\% of all gluino decays to 
  neutralinos for Model Lines {1} and {2}, these decays constitute about
  a third of all gluino to neutralino decays for Model Line {3}. This is, of
  course, due to the increased Yukawa coupling of the bottom quark
  \cite{dreesprl}.
\item In each of these cases, we see that decays of the gluino to the
  wino-like charginos and neutralinos have relatively small branching
  fractions despite their large $SU(2)$ gauge couplings. This is because
  decays mediated by lighter third generation squarks that have large
  Yukawa couplings dominate because these are dynamically as well as
  kinematically favoured.
\end{enumerate}

The main message of this Table is that in the WMAP favoured HB/FP
regions of the mSUGRA model, decays to third generation quarks dominate
gluino decays. Moreover, although detailed decay patterns depend on both
$\tan\beta$ and the value of $\mu/M_1$ (\textit{ i.e.} on how deep we are in
the HB/FP region), the total branching fraction for decays to third generation
quarks is relatively insensitive to these details. Motivated by these
observations we begin our examination of the inclusive $b$-jet signal
for each of the model lines introduced above.

The major SM backgrounds to the multijet plus $\eslt$ signal, with or
without $b$-jets, come from $W \ {\rm or} \ Z+{\rm jet}$ production, from
$t\bar{t}$ production and from QCD production of light quarks and
gluons. In the last case, the $\eslt$ arises from neutrinos from $c$ and
$b$ quarks, from showering of $W, Z$ bosons and their subsequent decays
to neutrinos, and from energy
mismeasurement.  The backgrounds from these sources are shown in
Table~\ref{tab:back} for two representative choices of cuts discussed
below. Here $S_T$ is the transverse sphericity, $m_{\rm eff}$ is the scalar
sum of the $E_T$ of the four hardest jets and $\eslt$, $\Delta\phi$ is
the azimuthal angle between $\eslt$ and the hardest jet, and
$\Delta\phi_b$ is the azimuthal opening angle between the two hardest $b$
jets in events with $N_b\ge 2$.  Since we do not require multileptons in
our analysis, backgrounds from $WW, WZ$ and $ZZ$ as well as three vector
bosons processes are expected to be much smaller \cite{bcpt}.

\begin{table}[tb]
\begin{center}
\caption{Cross sections for the dominant SM backgrounds to SUSY
  processes at the CERN LHC for two representative choices of cuts,
  together with signal cross sections for three cases along the model lines
  introduced in the text.}
\bigskip
\begin{tabular*}{\textwidth}{@{\extracolsep{\fill}}lccc}
\hline
\hline
\multicolumn{1}{c}{Variables}&\multicolumn{3}{c}{Cut 1 (cut 2)}\\
\hline 
$N_j \ge$              & & 4    (5)           \\
$\eslt$ (GeV)  $>$     & & 400  (500)           \\
$E_T^{j1}$ (GeV) $>$   & & 400  (400)           \\
$E_T^{j2}$ (GeV) $>$   & & 250  (250)           \\
$E_T^{j3}$  (GeV) $>$  & & 150  (175)           \\
$E_T^{j4}$  (GeV) $>$  & & 100  (125)           \\
$m_{\rm eff}$ (GeV) $>$& & 1500 (2250)         \\
$S_T > $               & & 0.0  (0.0)             \\
$\Delta \phi <$        & & 180$^\circ$ (140$^\circ$)      \\
$\Delta \phi_b <$      & & 180$^\circ$ (180$^\circ$)     \\
$N_b \ge$        & $ 0 $ &   $ 1 $ &   $ 2 $  \\
\hline
\multicolumn{1}{c}{Source}&\multicolumn{3}{c}{$\sigma$ (fb)}\\
\hline 
QCD        &$\;\;$ 27.71  (0.636) $\;\;$ &$\;\;$  8.95  (0.112) $\;\;$&  
$\;\;$ 1.72  (0.009) $\;\;$ \\
$t\bar{t}$ &  5.17  (0.043) &  3.42  (0.034) & 1.10 (0.005)  \\
$W+{\rm jets}$  &  13.60 (0.462) &  1.42  (0.045) &  0.13 (0.002)   \\
$Z+{\rm jets}$  &  5.68  (0.180)  & 0.65  (0.018) &  0.05 (0.001)  \\
Total & 52.16 (1.32) & 14.44 (0.209) & 3.00 (0.017)\\ \hline
Case 1: $m_{\tg} = 1054$ GeV  & 12.1 (0.667)  & 9.25 (0.537) & 4.58 (0.290) \\
Case 2: $m_{\tg}=  1436$ GeV  & 3.81 (0.541) & 3.10 (0.47) & 1.68 (0.24) \\
Case 3: $m_{\tg} = 1705$ GeV  & 1.11 (0.230)  & 0.90 (0.185) & 0.50 (0.106) \\
\hline
\hline
\end{tabular*}
\label{tab:back}
\end{center}
\end{table}

We see that for both sets of cuts the SM background is dominated by
QCD. This differs from earlier results where it is argued that the QCD
background can be reduced to negligible levels by requiring the signal
to be sufficiently hard. Indeed, to track the reason for this
difference, we have done a very high statistics analysis of the QCD
background that was not possible in Ref.~\cite{bcpt}. Specifically, we
took particular care to divide the event generation into a large number
(50 bins for QCD, fewer for other backgrounds) of finely spaced hard
scattering $p_T$ bins, especially for lower values of the hard
scattering $p_T$ where the event weights are very large, even if the
efficiency for passing the cuts is low.\footnote{If, for any set of
cuts, we find no events in our simulation of a particular background, we
set this background cross section to a value corresponding to the single
event level in the smallest weight bin in our simulation.}  We have
checked that our backgrounds levels from $W+{\rm jets}$, $Z+{\rm jets}$
 and $t\bar{t}$
processes are in agreement with those in Ref.~\cite{bcpt} and attribute
the difference in the QCD background to statistics of the background
simulation.\footnote{In passing, we mention that we also found a
significant QCD contribution to the background in the multijets +
$1\ell$ + $\eslt$ channel. This is in contrast to the result in
Ref.~\cite{tovey} which was obtained using PYTHIA.  We attribute this
difference to the fact that PYTHIA does not include showering of $W$ and
$Z$ bosons in QCD events. While this may result in some double counting
of ``Drell-Yan'' $W$ and $Z$ production, we note that showering of
vector bosons from the final state quarks will, in general tend to
populate a different region of phase space.}  Since QCD typically
contributes about half the background in Table~\ref{tab:back}, we may
expect a small degradation of the LHC reach from these earlier
projections.\footnote{A more significant background issue may be that
the showering algorithms appear to obtain a significantly softer
distribution of the variable $m_{\rm eff}$ (introduced below) relative
to evaluations using exact multijet plus $W, Z$ production matrix
elements \cite{mangano}. We have nothing to say about this, except that
requiring additional $b$'s (and in the case of $W$, also requiring a
transverse mass cut) will reduce this background considerably. At the
very least, our analysis will indicate the extent to which the presence
of tagged $b$'s increases the LHC reach. We may add that the $E_T^c$
analysis of Ref.\cite{bcpt} that requires $\eslt$ together with just two
(rather than four) additional hard jets, with just one of these jets
from showering, may be more robust to these matrix element corrections.}

Also shown in Table~\ref{tab:back} is the SUSY signal for three points
along model lines. In our analysis, we consider a signal to be
observable with a given integrated luminosity if, (\textit{a})~its
statistical significance $N_S/\sqrt{N_B} \ge 5$, (\textit{b})~$N_S/N_B \ge
0.25$, and (\textit{c})~$N_S\ge 10$.  We see that Case 1 with
$m_{\tg}\simeq 1$~TeV which by early analyses should be easily
observable at the LHC is also observable using cuts 1 with an integrated
luminosity of just 10~fb$^{-1}$, and moreover, the significance of the
signal improves with increasing $b$ multiplicity. For Case 2, the SM
background with the softer cuts 1 is too large except in the $2b$
channel, where the signal is observable for an integrated luminosity $\ge
26$~fb$^{-1}$. However with the harder cuts of set 2 the signal,
though unobservable without $b$-tagging, should be observable in the $1b$
($2b$) channel for an integrated luminosity exceeding 26~fb$^{-1}$
(41~fb$^{-1}$). For Case 3, both $b$-tagging capability and an
integrated luminosity of at least 72~fb$^{-1}$ are essential for the
observability of the signal. It is clear that $b$-tagging improves the
reach of the LHC for points in the HB/FP region.

To quantify the improvement $b$-tagging makes to the capabilities of the
LHC for the detection of SUSY, we have re-computed the reach of the LHC
for each of the three model lines introduced above. Towards this end,
for every mSUGRA parameter point that we examined, we generated a set of
SUSY events using ISAJET. We also generated large samples of SM
background events. We passed these events through the toy calorimeter
simulation mentioned previously, and then analysed both the signal and
the background for the entire set of cuts ($5\times 5\times 6\times
5\times 3^4= 60750$ choices in all) listed in
Table~\ref{tab:cuts}. We regard the signal as observable if it satisfies
the observability criteria (\textit{a})--(\textit{c}) listed above for at
least one choice of cuts.

\begin{table}[tb]
\begin{center}
\caption{The set of cuts used to optimize the SUSY signal at the LHC
  along the hyperbolic branch of the mSUGRA model.  }
\bigskip
\begin{tabular*}{\textwidth}{@{\extracolsep{\fill}}lc}
\hline
\hline
\multicolumn{1}{c}{Variable}&\multicolumn{1}{c}{Values}\\
\hline
$N_j \ge$              & 4, 5, 6, 7, 8   \\
\hline
$\eslt$ (GeV)  $>$     & 400, 500, 600, 700, 800 \\
\hline
($E_T^{j1}$, $E_T^{j2}$, $E_T^{j3}$, $E_T^{j4}$) (GeV) $>$
& (400, 250, 150, 100),   \\
& (400, 250, 175, 125),   \\
& (400, 250, 200, 150),   \\
& (500, 350, 150, 100),   \\
& (500, 350, 175, 125),   \\
& (500, 350, 200, 150)   \\
\hline
$m_{\rm eff}$ (GeV) $>$ & $[ \eslt + \sum_{i=1}^4 E_T^{ji} ]_{\rm
min} +
200 \times n, n = 0, 1, 2, 3, 4$ \\
\hline
$S_T > $               & 0.0, 0.1, 0.2             \\ \hline
$\Delta \phi <$        & 180$^\circ$, 160$^\circ$, 140$^\circ$ \\
\hline
$\Delta \phi_b <$      & 180$^\circ$, 160$^\circ$, 140$^\circ$ \\
\hline
$N_b \ge$              & 0, 1, 2 \\
\hline
\hline
\end{tabular*}
\label{tab:cuts}
\end{center}
\end{table}

Notice that the cuts in this Table are much harder than those used in a
recent analysis of the LHC SUSY signal \cite{sasha}. This is because,
unlike in Ref.~\cite{sasha} where the cuts were designed to extract the
signal for a wide range of squark and gluino masses, here we focus
on the optimization of the signal in the portion of the HB/FP
region with heavy gluinos where the previous strategy fails.  Note also
that cut 1 in Table~\ref{tab:back} corresponds to the softest of these
cuts. Since, as we saw earlier, the signal for Case 1 (with
$m_{\tg} \simeq 1$~TeV) is comfortably observable with the present set
of cuts as well as those in Ref.~\cite{sasha}, there is no danger that
there will be a gap in the HB/FP region of parameter space where the
signal is unobservable with either strategy.

\begin{figure}
\postscript{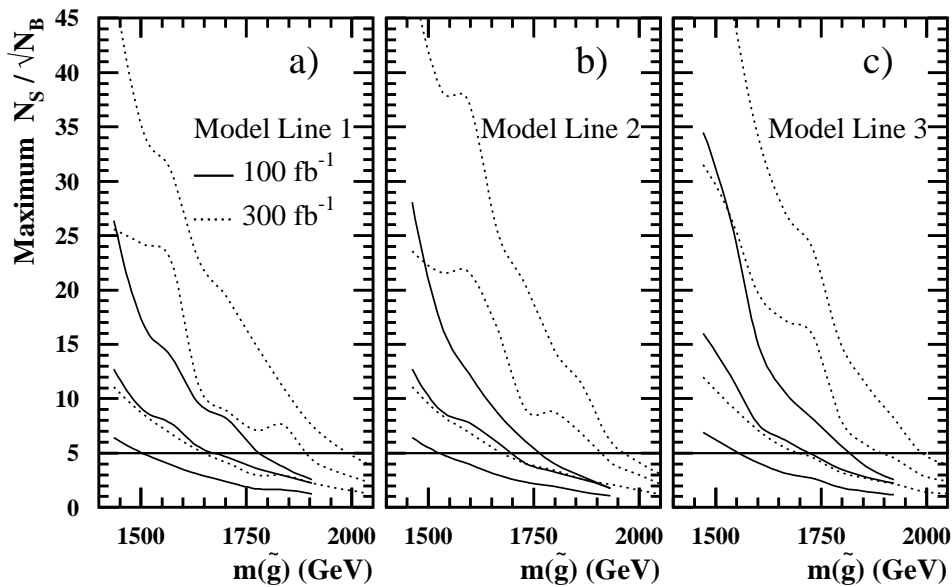}{0.95}
\caption{The maximum value of the statistical significance for the
  multijet + $\eslt$ SUSY signal at the LHC as we run over the cuts listed in
  Table~\ref{tab:cuts} for the three model lines introduced in the
  text, requiring in addition that $N_S \ge 10$ events and that $N_S \ge
  0.25N_B$. The solid lines from bottom to top denote this statistical
  significance without any $b$ tagging requirement, with at least one
  tagged $b$-jet, and with at least two tagged $b$-jets respectively,
  assuming an integrated luminosity of 100~fb$^{-1}$ and a tagging
  efficiency of 50\%. The dotted lines
  show the corresponding result for an integrated luminosity of
  300~fb$^{-1}$. 
\label{result}}
\end{figure}

Our results for the LHC reach are shown in Fig.~\ref{result}, where we
plot the largest statistical significance of the signal as we run over
the cuts in Table~\ref{tab:cuts} for a)~Model Line {1}, b)~Model
Line {2}, and c)~Model Line {3}. The solid curves, from lowest
to highest, denote this maximum statistical significance without
any $b$ tagging requirement, requiring $\ge 1$ tagged $b$-jet, and 
$\ge 2$ tagged $b$-jets respectively, for an integrated
luminosity of 100~fb$^{-1}$, while the dotted curves show the
corresponding results for 300~fb$^{-1}$ of integrated luminosity that
may be expected from three years of LHC operation with the high design
luminosity.  While, without $b$-tagging, our reach for gluinos is
$\sim 200$~GeV smaller than earlier projections \cite{sasha,atlas,cms},
presumably because of differences in the background levels discussed
above, we see that $b$-tagging will improve the mass reach of gluinos by
15-20\%, provided that LHC experiments can accumulate an integrated
luminosity of 100-300~fb$^{-1}$ and that $b$-tagging with an efficiency
of $\sim 50$\% remains possible even in the high luminosity environment.

A few remarks appear to be in order at this point: 
\begin{itemize}
\item The statistical significance in Fig.~\ref{result} is not
  significantly improved if the $b$-tagging efficiency improves to
  60\%. The reason is that before tagging the signal typically contains
  (on average) 3-4 $b$ quark jets while the background typically
  contains (at most) just two $b$ quark jets. As a result, the increased
  efficiency enhances the $b$-tagged background more than the signal,
  and the statistical significance is essentially unchanged. 

\item   We have
  also checked that with a $b$-tagging efficiency of 50\% and an
  integrated luminosity of 100~fb$^{-1}$, the signal with
  $\ge 3$ tagged $b$-jets is {\it rate limited} and no increase in the reach
  is obtained from that shown in the Figure.\footnote{It is possible
  that a slightly increased reach may be obtained if the tagging efficiency is
  significantly larger than 50\% or if the integrated luminosity is
  considerably higher than 100~fb$^{-1}$. It would, however, be
  necessary to evaluate backgrounds from $4b$, $4t$ and
  $t\bar{t}b\bar{b}$ production processes before a definitive conclusion
  can be made.}

\item The search strategy proposed here does not use lepton information
  at all. We have checked that a transverse mass cut $M_T(\ell, \eslt)
  \agt 100$~GeV on events with at least one isolated lepton does not
  increase the significance of the signal because the fraction of signal
  events (after our cuts) with an isolated lepton is not especially large. 

\item Although we have not shown this explicitly, we have checked that
  requiring the presence of additional isolated leptons does not lead to
  an increase in the reach relative to the $\ge 2b$ channel. 

\end{itemize}
  
In summary, we have shown that in the HB/FP region of the mSUGRA model
the reach of the LHC as measured in terms of gluino masses may be
increased by 15-20\% by requiring the presence of hard, tagged $b$-jets
in SUSY events. While we were mainly motivated in our investigation by
the fact that this part of parameter space is one of the regions
compatible with the WMAP data, our considerations may have wider
applicability since decays of heavy gluinos to third generation fermions
are favoured in all models with common masses for sfermions with the
same gauge quantum numbers. This is in part because the large top Yukawa
coupling and, if $\tan\beta$ is large, also the bottom quark Yukawa
coupling, cause the third generation squarks to be lighter than their
siblings in the first two generations, and in part because of new
contributions to gluino decay amplitudes from these large Yukawa
couplings \cite{dreesprl}.  Specifically, we may expect that signals
with tagged $b$-jets may also be useful in models with an inverted
squark mass hierarchy \cite{imh}, in models with unification of Yukawa
couplings (because they require large $\tan\beta$), and possibly also in
models with non-universal Higgs mass parameters that have recently been
re-examined in light of the WMAP data \cite{ournuhm}. Since the CMS and
ATLAS experiments are expected to ultimately have good $b$-tagging
capability, we urge that it be utilised to maximize the SUSY reach of
the LHC.

\acknowledgments
We thank F.~Gianotto and F.~Paige for discussions about $b$ tagging at
the LHC, and H.~Baer and M.~Drees for comments on the manuscript. 
This research was supported in part by the U.~S. Department of Energy
under contract number DE-FG-03-94ER40833 and by Funda\c{c}\~{a}o de Amparo 
\`a  Pesquisa do Estado de S\~ao Paulo (FAPESP).

%
%

%

\end{document}